\documentclass[10pt,letterpaper,twocolumn]{article}
\usepackage{amsfonts} 

\usepackage{ol2}
\usepackage[draft]{hyperref}
\usepackage{amsmath}
\usepackage{amssymb}

\begin{document}

\twocolumn[

\title{The Physics of Compressive Sensing and the Gradient-Based Recovery Algorithms }

\author{Qi~Dai and Wei~Sha}

\address{
Department of Electrical and Electronic Engineering, The University
of Hong Kong, Hong Kong, China.\\Email: daiqi@hku.hk (Qi~Dai);
wsha@eee.hku.hk (Wei Sha)\\\bf{Research Report}}
\begin{abstract}
\indent The physics of compressive sensing (CS) and the
gradient-based recovery algorithms are presented. First, the
different forms for CS are summarized. Second, the physical meanings
of coherence and measurement are given. Third, the gradient-based
recovery algorithms and their geometry explanations are provided.
Finally, we conclude the report and give some suggestion for future
work.

\end{abstract}
]

\section{Introduction}

\indent \indent The well-known Nyquist/Shannon sampling theorem that
the sampling rate must be at least twice the maximum frequency of
the signal is a golden rule used in visual and audio electronics,
medical imaging devices, radio receivers and so on. However, can we
simply recover a signal from a small number of linear measurements?
Yes, we can, answered firmly by Emmanuel J. Cand\`{e}s, Justin
Romberg, and Terence Tao \cite{1}\cite{2}\cite{3}. They brought us
the tool called Compressive Sensing (CS) \cite{4}\cite{5}\cite{6}
several years ago which avoids large digital data set and enables us
to build the data compression directly from the acquisition. The
mathematical theory underlying CS is deep and beautiful and draws
from diverse fields, but we don't focus too much on the mathematical
proofs. Here, we will give some physical explanations and discuss
relevant recovery algorithms.

\section{Exact Recovery of Sparse Signals}

\indent \indent Given a time-domain signal $f\in{\mathbb{R}^{N\times
1}}$, there are four different forms for CS. (a) If $f$ is sparse in
the time-domain and the measurements are acquired in the time-domain
also, then the optimization problem can be given by
\begin{align}\label{1}
\min{\|f\|_{{1}}}\,\,\,\,s.t.\,\,\,\,\mathcal{M}_{0}f=y
\end{align}
where $\mathcal{M}_{0}\in{\mathbb{R}^{M\times N}}$ is the
observation matrix and $y\in{\mathbb{R}^{M\times 1}}$ are the
measurements. (b) If $f$ is sparse in the time-domain and the
measurements are acquired in the transform-domain (Fourier
transform, discrete cosine transform, wavelet transform, X-let
transform, etc), then the optimization problem can be given by
\begin{align}\label{2}
\min{\|\Psi^{\dag}\tilde{f}\|_{{1}}}\,\,\,\,s.t.\,\,\,\,
\mathcal{M}_{0}\tilde{f}=\tilde{y}
\end{align}
where $\Psi^{\dag}$ is the inverse transform matrix and satisfies
$\Psi^{\dag}\Psi=\Psi\Psi^{\dag}=I$. (c) If $f$ is sparse in the
transform-domain and the measurements are acquired in the
time-domain, then the optimization problem can be given by
\begin{align}\label{3}
\min{\|\Psi f\|_{{1}}}\,\,\,\,s.t.\,\,\,\,\mathcal{M}_{0}f=y.
\end{align}
(d) If $f$ is sparse in the transform-domain and the measurements
are acquired in the transform-domain also, then the optimization
problem can be given by
\begin{align}\label{4}
\min{\|\tilde{f}\|_{{1}}}\,\,\,\,s.t.\,\,\,\,
\mathcal{M}_{0}\tilde{f}=\tilde{y}.
\end{align}

From the above equations, the meanings of the sparsity can be
generalized. If the number of the non-zero elements is very small
compared with the length of the time-domain signal, the signal is
sparse in the time-domain. If the most important $K$ components in
the transform-domain can represent signal accurately, we can say the
signal is sparse in the transform-domain. Because we can set other
unimportant components to be zero and implement the inverse
transform, the time-domain signal can be reconstructed with very
small numerical error. The sparsity property also makes the lossy
data compression possible. For the image processing, the derivatives
of the image (especially for the geometric image) along the
horizontal and vertical directions are sparse. For the physical
society, we can say the wave function is sparse with the specific
basis representations. Before you go into the CS world, you must
know what is sparse in what domain.

The second question is what is the size limit for the measurements
$y$ in order to perfectly recover the $K$ sparse signal. Usually,
$M\gtrsim K\log_2(N)$ or $M\approx4K$ for the general signal or
image. Further, if the signal $f$ is sparse in the transform-domain
$\Psi$ and the measurements are acquired in the time-domain, then
$M\gtrsim\chi\left(\Psi,\mathcal{M}_{0}\right)K\log_2(N)$, where
$\chi\left(\Psi,\mathcal{M}_{0}\right)$ is the coherence index
between the basis system $\Psi$ and the measurement system
$\mathcal{M}_{0}$ \cite{3}. The incoherence leads to the small
$\chi$ and therefore fewer measurements are required. The coherence
index $\chi$ can be easily found, if we rewrite \eqref{3} as
\begin{align}\label{5}
\min{\|\tilde{f}\|_{{1}}}\,\,\,\,s.t.\,\,\,\,\mathcal{M}_{0}\Psi^{\dag}\tilde{f}=y.
\end{align}
Similarly, \eqref{2} can be rewritten as
\begin{align}\label{6}
\min{\|f\|_{{1}}}\,\,\,\,s.t.\,\,\,\, \mathcal{M}_{0}\Psi
f=\tilde{y}
\end{align}

Third, what are the inherent properties for the observation matrix?
The observation matrix obeys what is known as a uniform uncertainty
principle (UUP).
\begin{align}
C_1\frac{M}{N}\leq\frac{\left|\left|\mathcal{M}_{0}f\right|\right|_2^2}
{\left|\left|f\right|\right|_2^2}\leq C_2\frac{M}{N}
\end{align}
where $C_1\lesssim1\lesssim C_2$. An alternative condition, which is
called restricted isometry property (RIP), can be given by
\begin{align}
1-\delta_k\leq\frac{\left|\left|\mathcal{M}_{0}f\right|\right|_2^2}
{\left|\left|f\right|\right|_2^2}\leq 1+\delta_k
\end{align}
where $\delta_k$ is a constant and is not too close to $1$. The
properties show the three facts: (a) The measurements $y$ can
maintain the energy of the original time-domain signal $f$. In other
words, the measurement process is stable. (b) If $f$ is sparse, then
$\mathcal{M}_{0}$ must be dense. This is the reason why the theorem
is called UUP. (c) If we want to perfectly recover $f$ from the
measurements $y$, at least $2K$ measurements are required. According
to the UUP and RIP theorems, it is convenient to set the observation
matrix $\mathcal{M}_{0}$ to a random matrix (normal distribution,
uniform distribution, or Bernoulli distribution).

\begin{figure}[htb]
\centering \centerline{\includegraphics[width=7
 cm]{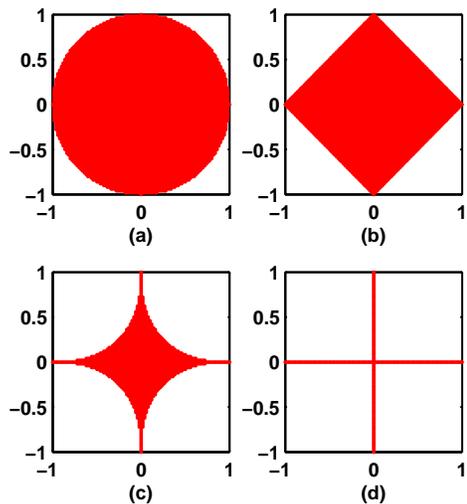}} \caption{The geometries of $l_p$ norm: (a) $p=2$;
(b) $p=1$; (c) $p=0.5$; (d) $p=0$.}\label{Fig1}
\end{figure}

Four, why $l_1$ norm is used in
\eqref{1}\eqref{2}\eqref{3}\eqref{4}? For the real application, the
size of measurement $M\ll N$. As a result, one will face the problem
how to solve an underdetermined matrix equation. In other words,
there are a huge amount of different candidate signals that could
all result in the given measurements. Thus, one must introduce some
additional constraints to select the ¡°best¡± candidate. The
classical solution to such problems would be minimizing the $l_2$
norm (the pseudo-inverse solution), which minimizes the amount of
energy in the system. However, this leads to poor results for most
practical applications, as the recovered signal seldom has zero
components. A more attractive solution would be minimizing the $l_0$
norm, or equivalently maximize the number of zero components in the
basis system. However, this is NP-hard (it contains the subset-sum
problem), and so is computationally infeasible for all but the
tiniest data sets. Thus, the $l_1$ norm, or the sum of the absolute
values, is usually what is to be minimized. Finding the candidate
with the smallest $l_1$ norm can be expressed relatively easily as a
linear convex optimization program, for which efficient solution
methods already exist. This leads to comparable results as using the
$l_0$ norm, often yielding results with many components being zero.
For simplicity, we take the 2-D case for example. The boundaries of
$l_0$, $l_1$, and $l_2$ norms are cross, diamond, and circle,
respectively. (See Fig.~\ref{Fig1}). The underdetermined matrix
equation can be seen as a straight line. If the intersection between
the straight line and the boundary is located at the x-axis or
y-axis, the recovered result will be sparse. Obviously, the
intersection will always be located at the axes if ${p}\leq1$ for
the $l_p$ norm.

Five, can CS have a good performance in a noisy environment? Yes, it
can. Because the recovery algorithm can get the most important $K$
components and force other components (including noise components)
to be zero. For the image processing, the recovery algorithm will
not smooth the image but yield the sharp edge.

Finally, let us review how CS encode and decode the time-domain
signal $f$. For the encoder, it gets the measurements $y$ or
$\tilde{y}$ according to the observation matrix $\mathcal{M}_{0}$.
For the decoder, it recovers $f$ or $\tilde{f}$ by solving the
convex optimization problem \eqref{1}\eqref{2}\eqref{3}\eqref{4}
with $y$ ($\tilde{y}$), $\mathcal{M}_{0}$, and $\Psi$\footnote{For
\eqref{1}\eqref{4}, $\Psi$ is not necessary.}. Hence, CS can be seen
as a fast encoder with lower sampling rate (fewer data set). The
sampling rate only depends on the sparsity of $f$ in some domains
and goes beyond the limit of the Nyquist/Shannon sampling theorem.
However, CS will face a challenging problem: How to perfectly
recover the signal with low computational complexity and memory?

\section{Physics of Compressive Sensing}
\indent \indent The most important concepts of the CS theory involve
Coherence and Measurement.

In physics, coherence is a property of waves, that enables
stationary (i.e. temporally and spatially constant) interference.
More generally, coherence describes all correlation properties
between physical quantities of a wave. When interfering, two waves
can add together to create a larger wave (constructive interference)
or subtract from each other to create a smaller wave (destructive
interference), depending on their relative phase. The coherence of
two waves follows from how well correlated the waves are as
quantified by the cross-correlation function.  The cross-correlation
quantifies the ability to predict the value of the second wave by
knowing the value of the first. As an example, consider two waves
perfectly correlated for all times. At any time, if the first wave
changes, the second will change in the same way. If combined they
can exhibit complete constructive interference at all times. It
follows that they are perfectly coherent. So, the second wave needs
not be a separate entity. It could be the first wave at a different
time or position. In this case, sometimes called self-coherence, the
measure of correlation is the autocorrelation function. Take the
Thomas Young's double-slit experiment for example, a coherent light
source illuminates a thin plate with two parallel slits cut in it,
and the light passing through the slits strikes a screen behind
them. The wave nature of light causes the light waves passing
through both slits to interfere, creating an interference pattern of
bright and dark bands on the screen. In fact, the dark bands can
relate to the zero components in the signal processing field. It is
well known that we select the basis functions coherent with the
signal or image. If the signal is the square wave, the haar wavelet
is a good choice. If the signal is the sine wave, the Fourier
transform is a good choice. The coherence index between a signal and
a basis system will decide the sparsity of the signal in the
transform-domain. In other words, fewer basis functions will be used
or more components in the transform-domain are to be zero if the
signal and the basis functions are coherent. For the CS, however,
the observation matrix $\mathcal{M}_{0}$ and the time-domain signal
$f$ should be incoherent. In addition, the observation matrix
$\mathcal{M}_{0}$ and the basis system $\Psi$ also should be
incoherent. If it is not the case, the reconstruction matrix
$\mathcal{M}_{0}\Psi^{\dag}$ in \eqref{5} will be sparse, which
violates the UUP theorem.

Then, let us talk about quantum mechanics and the measurement. In
physics, a wave function or wavefunction is a mathematical tool used
in quantum mechanics to describe any physical system. The values of
the wave function are probability amplitudes (complex numbers). The
squares of the absolute values of the wave functions $|f|^2$ give
the probability distribution (the chance of finding the subject at a
certain time and position) that the system will be in any of the
possible quantum states. The modern usage of the term wave function
refers to a complex vector or function, i.e. an element in a complex
Hilbert space. An element of a vector space can be expressed in
different bases; and so the same applies to wave functions. The
components of a wave function describing the same physical state
take different complex values depending on the basis being used;
however the wave function itself is not dependent on the basis
chosen. Similarly, in the signal processing field, we use different
basis functions to represent the signal or image.

The quantum state of a system is a mathematical object that fully
describes the quantum system. Once the quantum state has been
prepared, some aspect of it is measured (for example, its position
or energy). It is a postulate of quantum mechanics that all
measurements have an associated operator\footnote{For the discrete
system, an operator can be seen as a matrix.} (called an observable
operator). The expected result of the measurement is in general
described not by a single number, but by a probability distribution
that specifies the likelihoods that the various possible results
will be obtained. The measurement process is often said to be random
and indeterministic. Suppose we take a measurement corresponding to
observable operator $\hat{O}$, on a state whose quantum state is
$f$. The mean value (expectation value) of the measurement is
$\langle f,\hat{O}f\rangle$ and the variance of the measurement is
$\langle f,\hat{O}^2f\rangle-(\langle f,\hat{O}f\rangle)^2$. Each
descriptor (mean, variance, etc) of the measurement involves a part
of information of the quantum state. In the signal processing field,
the observation matrix $\mathcal{M}_{0}$ is a random matrix and each
measurement $y_i$ captures a portion of information of the signal
$f$. Due to the UUP and RIP theorems, all the measurements make the
same contribution to recovering $f$. In other words, each
measurement $y_i$ is equally important or unimportant\footnote{For
the traditional compression method, the perfect reconstruction is
impossible if some important components are lost.}. The unique
property will make CS very powerful in the communication field
(channel coding).

\section{Gradient-Based Recovery Algorithms}
\indent\indent A fast, low-consumed, and reliable recovery algorithm
is the core of the CS theory. There are a lot of outstanding work on
the topic\cite{7}\cite{8}\cite{9}\cite{10}. Based on their work, we
developed the gradient-based recovery algorithms. In particular, we
did not reshape the image (matrix) to the signal (vector), which
will consume a large amount of memory. We treat each column of the
image as a vector and the comparable results also can be obtained.
For the sparse image in the time-domain, the $l_1$ norm constraint
is used. For the general image (especially for the geometric image),
the total variation constraint is used. Considering the
non-differentiability of the function $|f_{j,k}|$ at the origin
point, the subgradient or smooth approximation strategies\cite{10}
are employed.

\subsection{Gradient Algorithms and Their Geometries}
\indent\indent Before solving the constrained convex optimization
problems, the clear and deep understandings for the gradient-based
algorithms are necessary. Given a linear matrix equation
\begin{align}\label{7}
\mathcal{M}_{0}f=y
\end{align}
the solution $f$ can be found by solving the following minimization
problem
\begin{align}\label{8}
\min_{f} L(f)\equiv \frac{1}{2}||\mathcal{M}_{0}f-y||_{2}^{2}
\end{align}
The gradient-based algorithms for solving \eqref{8} can be written
as
\begin{align}
f^{i+1}=f^{i}-\mu^i\nabla L(f^{i})
\end{align}
where $\mu^i$ is the iteration step size and
\begin{align}
\nabla L(f^{i})=\mathcal{M}_{0}^{\dag}(\mathcal{M}_{0}f^{i}-y)
\end{align}
A variety of settings for $\mu^i$ result in different algorithms
involving the gradient method, the steepest descent method, and the
Newton's method. The gradient method sets $\mu^i$ to a small
constant. The steepest descent method sets $\mu^i$ to
\begin{align}\label{9}
\mu^i=\frac{\left\langle\nabla L(f^{i}),\nabla
L(f^{i})\right\rangle}{\left\langle\nabla
L(f^{i}),\mathcal{M}_{0}^{\dag}\mathcal{M}_{0}\nabla
L(f^{i})\right\rangle+\varepsilon}
\end{align}
which minimizes the residual $R^i=\mathcal{M}_{0}f^{i}-y$ in each
iteration. Here, the small $\varepsilon$ is used for avoiding a zero
denominator. For the Newton's method, $\mu^i$ is taken as a constant
matrix
\begin{align}\label{10}
\mu^i=\left(\mathcal{M}_{0}^{\dag}\mathcal{M}_{0}+\varepsilon
I\right)^{-1}
\end{align}
Here, the small $\varepsilon$ is used for avoiding a nearly singular
matrix.

\begin{figure}[htb]
\centering \centerline{\includegraphics[width=7 cm]{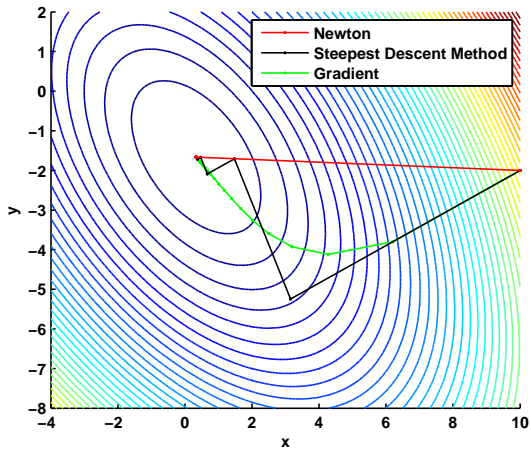}}
\caption{The geometries for the gradient-based
algorithms.}\label{Fig2}
\end{figure}

To understand the geometries for the three gradient-based
algorithms, a simple case is taken for example. A 2-D function
$f(x,y)=(x+y)^2+(x+1)^2+(y+3)^2$ has a local minimum
$f^{*}=\left(\frac{1}{3},-\frac{5}{3}\right)$. The contour of the
function and the trajectory of $f^i$ are drawn in Fig.~\ref{Fig2}.

\noindent The convergence of the gradient method is worst. The
steepest descent method converges fast at the first several steps
but slowly as the iteration step increases. The Newton's method is
best and needs only one step for the 2-D case\footnote{For the CS,
the performance of the Newton's method will decrease due to the
nearly singular observation matrix and the $l_1$ norm constraint.}.

Next, we will apply the steepest descent method and the Newton's
method to recover the signal or image $f$ by using \eqref{1}. The
treatments for other convex optimization problems
\eqref{2}\eqref{3}\eqref{4} are similar.

\subsection{$l_1$ Norm Strategy}

\indent\indent We assume $f_{j,k}$ is the pixel of an $N\times N$
image $f$ at the j-th row and the k-th column ($1\leq j\leq N$ and
$1\leq k\leq N$). The convex optimization problem \eqref{1} for the
sparse image can be converted to
\begin{align}
\min_{f} H(f)\equiv L(f)+\lambda \|f\|_{1}
\end{align}
where $\|f\|_{1}=\sum_{j,k}|f_{j,k}|$. The above equation is a
Lagrange multiplier formulation. The first term relates to the
underdetermined matrix equation \eqref{7} and the second
$l_1$-penalty term will assure a regularized sparse solution. The
parameter $\lambda$ balances the weight of the first term and the
second term.

Because $|f_{j,k}|$ is not differentiable at the origin point, we
can define a new subgradient for each $f_{j,k}$ as follows
\begin{align}
\nabla_{j,k}{H(f)}= \left \{
\begin{array}{cl}
  & \nabla_{j,k}{L(f)}+\lambda\,\mathrm{sign}(f_{j,k}),\,\,|f_{j,k}|\geq\varepsilon \\
  & \nabla_{j,k}L(f)+\lambda,\,\,|f_{j,k}|<\varepsilon,\,\,\nabla_{j,k}L(f)<-\lambda \\
  & \nabla_{j,k}L(f)-\lambda,\,\,|f_{j,k}|<\varepsilon,\,\,\nabla_{j,k}L(f)>\lambda \\
  &
0,\,\,|f_{j,k}|<\varepsilon,\,\,\left|\nabla_{j,k}L(f)\right|\leq\lambda
\end{array}
\right.
\end{align}
Then the gradient-based algorithm can be written as
\begin{align}\label{12}
f_{j,k}^{i+1}=f_{j,k}^{i}-\mu_k^i\nabla_{j,k}H(f^{i})
\end{align}
where $j$ and $k$ are, respectively, the row index and the column
index of the image $f$ and $i$ denotes the i-th iteration step. The
$\mu_k^i$ has been given in \eqref{9} and \eqref{10}. Bear in mind,
the image will be treated column by column when computing $\mu_k^i$
and $\nabla_{j,k}{L(f^i)}$.

For the steepest descent method, the parameter $\lambda$ can be
taken as a small positive constant ($\lambda=0.001-0.01$). But for
the Newton's method, the parameter $\lambda$ must be gradually
decreased as the iteration step increases
($\lambda^{i+1}=(0.99-0.999)\times\lambda^{i}$).

\subsection{Total Variation Strategy}

\indent\indent For a general image, especially for a geometric
image, it is not sparse in the time-domain. Hence, the $l_1$ norm
strategy developed in the previous subsection will break down.

The convex optimization problem for the general image can be given
by
\begin{align}
\min_{f} H(f)\equiv L(f)+\lambda\,\rm{TV}(f)
\end{align}
where $\rm{TV}(f)$ is the total variation of the image $f$. The
derivatives of $f$ along the vertical and horizontal directions can
be defined as
\begin{align}
D_{j,k}^{v}f=\left\{
\begin{array}{lc}
  f_{j,k}-f_{j+1,k} & 1\leq j<N \\
  0 & j=N
\end{array}
\right.
\end{align}
\begin{align}
D_{j,k}^{h}f=\left\{
\begin{array}{lc}
  f_{j,k}-f_{j,k+1} & 1\leq k<N \\
  0 & k=N
\end{array}
\right.
\end{align}
The total variation of the image $f$ is the summation for the
magnitude of the gradient of each pixel \cite{11}
\begin{align}\label{11}
\rm{TV}(f)=\sum_{j,k}{\sqrt{\left(D_{j,k}^{v}f\right)^2+\left(D_{j,k}^{h}f\right)^2}}
=\sum_{j,k}\left|\nabla_{j,k}f\right|
\end{align}
After some simple derivations, the gradient of the total variation
with each pixel is given by
\begin{align}\label{13}
\begin{split}
\nabla_{j,k}\left(\rm{TV}(f)\right)&=\frac{D_{j,k}^{v}f}{\left|\nabla_{j,k}f\right|}+
\frac{D_{j,k}^{h}f}{\left|\nabla_{j,k}f\right|}\\
&-\frac{D_{j-1,k}^{v}f}{\left|\nabla_{j-1,k}f\right|}
-\frac{D_{j,k-1}^{h}f}{\left|\nabla_{j,k-1}f\right|}
\end{split}
\end{align}
When treating \eqref{13}, the smooth approximation strategy is used
for avoiding a zero denominator, i.e.
\begin{align}
\left|\nabla_{j,k}f\right|=\sqrt{\left(D_{j,k}^{v}f\right)^2+\left(D_{j,k}^{h}f\right)^2+\varepsilon}
\end{align}
The gradient-based algorithm has been given in \eqref{12}.

\section{Numerical Experiments and Results }

\subsection{Cases of $l_1$ Norm Strategy}
\begin{figure}[htb]
\centerline{\includegraphics[width=5cm]{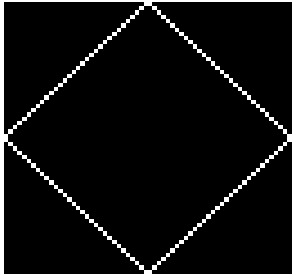}} \caption{The
$64\times64$ sparse diamond.}\label{Fig3}
\end{figure}

\indent\indent The first image we'd like to recover is a $64\times
64$ sparse diamond as shown in Fig.~\ref{Fig3}.

\begin{figure}[htb]
\centerline{\includegraphics[width=8
 cm]{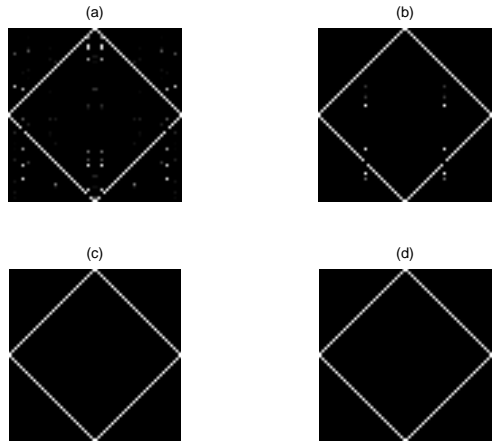}} \caption{The recovered diamond figures by the Newton's
method from the observation matrices with different sizes: (a)
$10\times 64$; (b) $12\times 64$; (c) $15\times 64$; (d) $20\times
64$.}\label{Fig4}
\end{figure}

\noindent Notice that the image itself is sparse in the time-domain,
we need not to transform the image into other domains, such as the
wavelet domain or Fourier domain. The size of the observation matrix
for the nearly perfect reconstruction should be at least larger than
$12\times 64$ where $12$ is calculated by $2\cdot \log_2{64}=12$. If
our observation matrices are generated by the uniform distribution
from $0$ to $1$, after $20000$ iteration steps, Fig.~\ref{Fig4} can
be obtained. The subplot (a) shows the recovered image with a
$10\times 64$ observation matrix, while (b), (c), and (d) are
recovered with $12\times 64$, $15\times 64$, and $20\times 64$
random observation matrices respectively. When the size of the
observation matrix is small, the poor reconstructed images are shown
in (a) and (b). But when the observation matrix gets larger and
larger, the better results can be obtained as shown in (c) and (d).

\begin{figure}[htb]
\centering\centerline{\includegraphics[width=5
 cm]{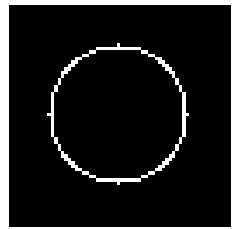}} \caption{The $64\times 64$ sparse circle.}\label{Fig5}
\end{figure}
\begin{figure}[htb]
\centerline{\includegraphics[width=8
 cm]{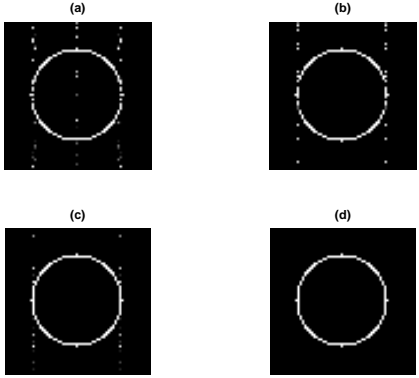}} \caption{The recovered circle figures by the steepest
descent method from the observation matrices with different sizes:
(a) $15\times 64$; (b) $20\times 64$; (c) $25\times 64$; (d)
$30\times 64$.}\label{Fig6}
\end{figure}

Another $64\times 64$ image we wish to recover is a circle as shown
in Fig.~\ref{Fig5}. Although the circle is sparse on the whole, but
it is not the case for some columns. These columns may require more
measurements if we treat an $n\times n$ image as $n$ vectors. Using
the steepest descent method, we can recover the image after several
iterations. Here, we don't want to compare the Newton's method with
the steepest descent method to find which one is more powerful and
effective. (Actually, their performances are almost the same for the
image.) What we want to show is the sparsity of an image affects the
recovered results a lot. Fig.~\ref{Fig6} shows the recovered results
from the observation matrices with different sizes slightly larger
than the previous case. For the subplots (a) and (b), the results
are undesirable. The subplot (c) is better and almost perfect
reconstruction is obtained for the subplot (d). Again, we notice
that for the small observation matrix, the recovered results may
vary drastically. If the size of the observation matrix is large
enough, the reconstructed image is accurate or even exact in any
repeated experiments. The subplots (a), (b), and (c) in
Fig.~\ref{Fig6} show that those columns which are less sparse are
the hardest to recover when the small observation matrix is
utilized. For the case, the total variation strategy may be a better
choice. In a word, a large observation matrix can capture more
information of the image and therefore the image can be recovered
with higher probability.

\subsection{Cases of Total Variation Strategy}
\begin{figure}[htb]
\centerline{\includegraphics[width=5
 cm]{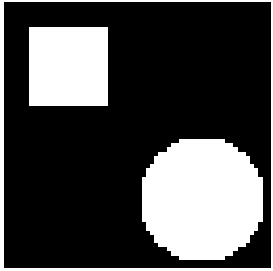}} \caption{The geometric figure.}\label{Fig7}
\end{figure}

\indent\indent If the image is not sparse in the time-domain, it
also can be recovered from the measurements without represented as
the basis functions $\Psi$ whose coefficients may be sparse. For
instance, the geometric figure composed of a solid circle and a
solid square is shown in Fig.~\ref{Fig7}. It is easy to imagine the
derivatives of the image are sparse. We apply the total variation
strategy to recover the image.

\begin{figure}[htb]
\centerline{\includegraphics[width=4
 cm]{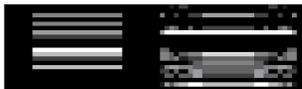}} \caption{The measurement of the geometric
figure.}\label{Fig8}
\end{figure}

\begin{figure}[htb]
\centerline{\includegraphics[width=5
 cm]{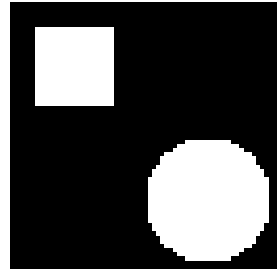}} \caption{The recovered geometric image by the steepest descent method: $20\times 64$
observation matrix is employed.}\label{Fig9}
\end{figure}

The size of the object image again is $64\times 64$ and the
$20\times 64$ observation matrix is employed. Fig.~\ref{Fig8} is the
typical measurements $y$ and Fig.~ \ref{Fig9} is the reconstructed
image from $y$. The peak signal to noise ratio (PSNR) calculated is
always above $90$ for the repeated experiments (different
observation matrices with the same size), which suggests that the
observation matrix is large enough to recover the original image
with an extremely accurate result.

\begin{figure}[htb]
\centerline{\includegraphics[width=9
 cm]{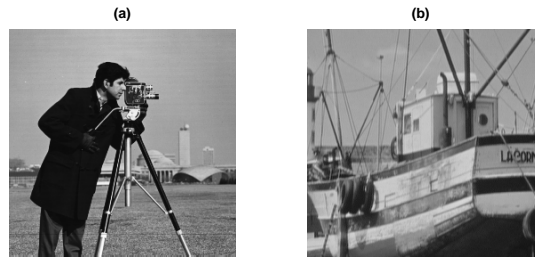}} \caption{The general image: (a) Cameraman; (b) Boats.}\label{Fig10}
\end{figure}

\begin{figure}[htb]
\centerline{\includegraphics[width=9
 cm]{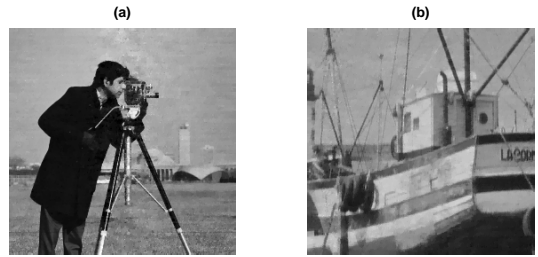}} \caption{The recovered images by the Newton's method:
$100\times 256$ observation matrix is employed. (a) Cameraman; (b)
Boats.}\label{Fig11}
\end{figure}

The next two images Cameraman and Boats in Fig.~\ref{Fig10} are
quite well-known in the image processing field. We still treat the
$256\times 256$ image as $256$ vectors. Although the result may not
be so good as that we obtain by treating the $256\times 256$ image
as a long vector of size $65536\times 1$, we do save a great amount
of memory and calculation time. The size of our observation matrix
$\mathcal{M}_{0}$ is $100\times 256$ rather than $25600\times 65536$
($25600\approx65536/2.56$). The recovery algorithm is implemented in
the wavelet-domain, and the PSNR for the subplots (a) and (b) in
Fig.~\ref{Fig11} are $29.4$ and $30.9$, respectively. In addition,
the gradient-based total variation algorithm also has a good
performance for the geometric image Peppers. (Showing too many
results are not necessary).

\section{Discussions and Future Work}
\indent\indent The above are just some simple experiments for
demonstrating that the CS was able to recover an image accurately
from a few of random projections. One should understand that the
main advantage of CS is not how small size it can compress the image
to. In fact, if a signal is $K$ sparse in some domains, we indeed
require $3K$ to $5K$ measurements to recover the signal. An obvious
advantage of CS is that it can encode the signal or image fast. In
particular, the prior knowledge about the signal is not important.
For example, it is not necessary for us to know the exact positions
and values of the most important components beforehand. What we care
is whether the image is sparse in some domains or not. A fixed
observation matrix can be applied to measure different signals,
which makes the applications of CS for encoding and decoding
possible. Meanwhile, the measurements play the same role in
recovering the signal or image, which makes CS very powerful in
military applications (radar imaging) where we cannot afford the
risk caused by the loss of the most important $K$ components. Since
each random projection (measurement) is equally (un)important, the
CS is not sensitive to the noises or measurement errors and can
provide the robust and stable performances.

Although many researchers have made great progresses in the convex
optimization problems and demonstrated the accurate results on the
scale we interest (hundreds of thousands of measurements and
millions of pixels), the more efficient algorithms are still
required. Actually, solving the $l_1$ minimization problem is about
30-50 times as expensive as solving the least squares problem.
However, the unbalanced computational burden gives us a chance that
the measurements are acquired by the sensors with lower power, and
then the signal or image will be recovered on the central
supercomputer. The algorithms, such as the conjugate gradient method
and the generalized minimal residual method, will become our next
candidates for accelerating the recovery algorithm.

The physical understandings and applications for CS are under way,
although a single-pixel camera has shocked the field of optics. We
are aware that the CS has penetrated many fields and become a
hotspot. We expect more mathematicians, physicist, and engineers
make contributions for the CS field.

\section{Acknowledgement}
\indent\indent The authors are not from signal or image processing
society. Hence, some technical words may not be right. We hope the
report can give some help for the researchers form the signal
processing, image processing, and physical societies.

\end{document}